\documentclass[10pt, conference]{IEEEtran} 
\usepackage{booktabs}
\usepackage{float}
\usepackage{pifont}
\usepackage{url}
\usepackage{hyperref}
\usepackage[hyphenbreaks]{breakurl}

\IEEEoverridecommandlockouts
\usepackage{cite}
\usepackage{amsmath,amssymb,amsfonts, mathtools}
\usepackage{algorithmic}
\usepackage{graphicx}
\usepackage{textcomp}
\usepackage{xcolor}
\usepackage{adjustbox}
\usepackage[ruled,vlined, linesnumbered]{algorithm2e}
\def\BibTeX{{\rm B\kern-.05em{\sc i\kern-.025em b}\kern-.08em
    T\kern-.1667em\lower.7ex\hbox{E}\kern-.125emX}}
\graphicspath{{images/}}

\usepackage{balance}

\usepackage{orcidlink}

\usepackage{todonotes}

\usepackage{listings, color}
\PassOptionsToPackage{linktocpage}{hyperref}
\hypersetup{
    breaklinks=true,   
    pdfusetitle=true,  
}

\lstdefinelanguage{QASM}
{
	morekeywords={
		if,
		gate,
		qreg,
		creg,
		measure,
		while,
		qubit,
		bit
	},
	sensitive=false, 
	morecomment=[l]{//}, 
	morecomment=[s]{/*}{*/}, 
	morestring=[b]"
}
\definecolor{Green}{RGB}{34,139,34}
\definecolor{Blue}{RGB}{83,140,144}
\definecolor{Red}{RGB}{189,55,57}

\usepackage[font=footnotesize,labelfont=bf]{caption}

\newcommand{\defineReviewCommands}[2]{
	\expandafter\def\csname#1\endcsname##1{\todo[color=#2]{##1 -#1}}
	\expandafter\def\csname#1i\endcsname##1{\todo[inline,caption={},color=#2]{##1 -#1}}
}

\defineReviewCommands{joshua}{cyan}
\defineReviewCommands{christoph}{red}
\defineReviewCommands{julian}{green}

\usepackage{pgfplots}
\pgfplotsset{compat=1.18}
\usepackage{tikz}
\usetikzlibrary{quantikz, fit, patterns, calc}
\tikzset{
  font={\small}
}

\usepackage{threeparttable}
\usepackage{tcolorbox}
\usepackage{enumitem}
\setlist[description]{labelindent=\parindent}

\usepackage{subcaption}
\usepackage{multirow}
\usepackage[bottom, hang]{footmisc}
\SetArgSty{textnormal}

\begin{document}

\makeatletter
\newcommand{\linebreakand}{%
  \end{@IEEEauthorhalign}
  \hfill\mbox{}\par
  \mbox{}\hfill\begin{@IEEEauthorhalign}
}
\makeatother

\title{Quantum Pattern Detection:\\Accurate State- and Circuit-based Analyses

\thanks{This work has been supported by the German Ministry for Education and Research in project QuBRA (reference number: 13N16303).}
}

\author{
    \IEEEauthorblockN{
         Julian Shen\IEEEauthorrefmark{1}\orcidlink{0009-0009-6591-0172}, 
         Joshua Ammermann\IEEEauthorrefmark{1}\orcidlink{0000-0001-5533-7274}, 
         Christoph König\IEEEauthorrefmark{1}\orcidlink{0009-0009-5945-1029}, 
         and Ina Schaefer\IEEEauthorrefmark{1}\orcidlink{0000-0002-7153-761X}
    }\\
    \linebreakand
    \IEEEauthorblockA{\IEEEauthorrefmark{1}
         Institute of Information Security and Dependability\\
         Karlsruhe Institute of Technology (KIT)\\
         Karlsruhe, Germany\\
         Email: julian.shen@student.kit.edu, \{name\}.\{surname\}@kit.edu
    }
 }

\maketitle


\begin{abstract}
Quantum computers have the potential to solve certain problems faster than classical computers by exploiting quantum mechanical effects such as superposition.
However, building high-quality quantum software is challenging due to the fundamental differences between quantum and traditional programming and the lack of abstraction mechanisms.
To mitigate this challenge, researchers have introduced quantum patterns to capture common high-level design solutions to recurring problems in quantum software engineering.
In order to utilize patterns as an abstraction level for implementation, a mapping between the theoretical patterns and the source code is required, which has only been addressed to a limited extent.
To close this gap, we propose a framework for the automatic detection of quantum patterns using state- and circuit-based code analysis.
Furthermore, we contribute a dataset for benchmarking quantum pattern detection approaches.
In an empirical evaluation, we show that our framework is able to detect quantum patterns very accurately and that it outperforms existing quantum pattern detection approaches in terms of detection accuracy.
\end{abstract}


\section{Introduction}
\label{sec:introduction}

Quantum computers possess the potential to outperform classical computers on certain computational problems by utilizing quantum physical effects like superposition and entanglement~\cite{Homeister.2018}.
However, building high-quality quantum software is challenging due to the fundamental differences between quantum and traditional programming and the absence of abstraction mechanisms. 
In contrast to classical algorithms which are executed sequentially on the program state, quantum algorithms consider multiple states at once and perform operations on them simultaneously~\cite{Homeister.2018}.
This makes the implementation of quantum algorithms a complex task since identifying principles that help construct reusable and maintainable quantum software can be difficult.

In classical software engineering, this challenge is mitigated through the documentation of design principles and best practices as \emph{patterns}~\cite{Beck.1987}.
Patterns provide a higher level of abstraction for implementation by capturing design and architectural knowledge in a human-readable format and act like pre-made blueprints for the construction of software systems~\cite{Ramasamy.2015}.
This makes patterns an indispensable concept for the understanding and development of larger software systems, as they facilitate programmers to think about implementation problems conceptually, simplify the coding process, and reduce the communication effort between software developers~\cite{Unger.2000}.

In the quantum computing domain, Leymann et al.~\cite{Leymann.2019} introduced a pattern language for the design of quantum algorithms which has been extended several times~\cite{Weigold.2021, Weigold.2022, Weigold.2021.2, Beisel.2022, Buehler.2023}.
The patterns are analogous to the pattern concept in classical computing and are grouped into different categories, including patterns for quantum operations~\cite{Leymann.2019}, data encoding~\cite{Weigold.2021, Weigold.2022}, hybrid quantum algorithms~\cite{Weigold.2021.2}, error handling~\cite{Beisel.2022}, and execution~\cite{Buehler.2023}.
However, these patterns were only documented as theoretical concepts, often without reference to concrete implementations on code level.
In order to make use of patterns as an abstraction level for quantum software development and to further increase the program understanding in the area of quantum computing, a mapping between the theoretical patterns and the actual source code is needed.
To create such a mapping, Pérez-Castillo et al.~\cite{Perez.2024} proposed a first approach for detecting five specific patterns in existing source code of quantum algorithms automatically in order to characterize the usage of patterns.
However, two of the five patterns were not found at all during the evaluation~\cite{Perez.2024}, and the number of patterns that can be found with their tool is extremely limited.
Thus, currently, there is no detection software that is capable of recognizing quantum computational patterns in a reliable manner.

To close this gap, we contribute a framework that analyzes different implementations of quantum algorithms by detecting eight patterns in the underlying program code automatically.
Our detection approaches include both static and dynamic code analysis and are evaluated against a benchmark dataset consisting of 20 quantum algorithms.
This data set can be later used for the evaluation of future pattern detection programs.
In summary, we make the following contributions:
\begin{itemize}
    \item We present two novel approaches for the detection of quantum patterns using static and dynamic code analysis.
    \item We contribute a dataset for benchmarking future pattern detection approaches.
    \item We demonstrate that our framework is able to detect patterns very accurately in an empirical evaluation and that it outperforms existing detection approaches for quantum computing patterns in terms of detection accuracy. 
\end{itemize}

\section{Motivation and Objectives}
\label{sec:motivation}
As software systems grow more complex, maintaining high-quality and manageable code becomes increasingly challenging~\cite{Fowler.2002}. 
A common strategy to reduce complexity is by introducing \emph{abstractions}, which distill the overwhelming details of software into high-level components. 
These abstractions allow developers to focus on key concepts, enabling systematic reasoning about large-scale software development.
In classical object-oriented programming, such abstractions have been established in the concept of \emph{patterns}~\cite{Beck.1987}.
At the highest level, architectural patterns~\cite{Shaw.1996} serve as templates for designing the coarse-grained structure of software systems.
These templates can then be filled with solutions to lower level problems using software design patterns~\cite{Gamma.1995}.
Therefore, patterns can be seen as abstract building blocks for constructing large-scale systems~\cite{Sinnig.2005} that provide off-the-shelf solutions to recurring problems.
In traditional programming, patterns have already been established as an essential part of high-quality software engineering~\cite{Buschmann.1996}.

In the quantum computing domain, which is dominated by physical and mathematical concepts, the notion of constructing software from predefined building blocks is particularly important.
Although Leymann et al.~\cite{Leymann.2019} have introduced a pattern language for solving typical problems in quantum computing, the language is currently not yet sufficient to build larger quantum systems from patterns alone.
It is also often unclear how the existing theoretical patterns can be concretely implemented.
Therefore, our primary objective is to conduct research towards the concept of using patterns as high-level building blocks in the field of quantum computing, i.e. the notion of constructing quantum software solely with patterns.
The framework that we contribute is another step towards this objective, as it provides an automatic analysis of quantum code and can help to gain a deeper understanding of quantum pattern usage.
In addition to that, it can be easily extended to help discover missing patterns by identifying code passages that are currently not covered by any pattern.
\section{Background}
\label{sec:background}
This section introduces fundamentals of quantum computing and provides explanations of the quantum computing patterns that can be detected using our framework.

\subsection{Quantum Computing}
Quantum computers perform calculations on \emph{qubits} which abstract the state of a quantum system and act as the fundamental unit of information.
The state of a qubit is represented as a linear combination of two orthonormal basis vectors that span a two-dimensional complex vector space.
In quantum computing, vectors are typically written in Dirac notation~\cite{Dirac.1939} where a vector $a$ is denoted inside a ket and represented as $\ket{a}$. 
Often, the vectors $\ket{0}=(1,0)^\top$ and $\ket{1}=(0,1)^\top$ are used as basis vectors and together, the set $\{\ket{0},\ket{1}\}$ is called the \emph{computational basis}~\cite{Nielsen.2011}.
Using these basis states, the state of a qubit $x$ can be expressed as $\ket{x}=\alpha\ket{0}+\beta\ket{1}$ where $\alpha,\beta\in\mathbb{C}$ are called \emph{probability amplitudes} satisfying the property $|\alpha|^2+|\beta|^2=1$.
In order to retrieve information about the state of a qubit, it has to be \emph{measured}.
Measurement collapses the superposition of a qubit and the result depends on the amplitudes $\alpha$ and $\beta$.
The probability of outcome $\ket{0}$ is $|\alpha|^2$ and the probability of outcome $\ket{1}$ is $|\beta|^2$.
Geometrically, the state of a qubit can be represented in a Bloch Sphere~\cite{Arecchi.1972}
where a possible state of the qubit is described by a point on the surface of the sphere.
 		

 		
 	\label{fig:sphere}
The state of qubits can be manipulated with \emph{quantum gates}.
Quantum gates can be divided into single-qubit gates and multi-qubit gates, depending on the number of qubits they act on.
One of the most common single-qubit gates is the Hadamard gate $H$ which moves a qubit from the state $\ket{0}$ into the \emph{uniform superposition} state $\frac{1}{\sqrt{2}}(\ket{0}+\ket{1})$~\cite{Leymann.2019} which means that all measurement outcomes of that qubit have equal probabilities.
A Hadamard gate can be combined with a controlled-NOT (CNOT) gate to \emph{entangle} the state of two qubits~\cite{Nielsen.2011},
shown in Fig.~\ref{fig:entanglement}.
\begin{figure}[tb]
	\centering
    \resizebox{0.6\linewidth}{!}{ 
	\begin{quantikz}[slice all, remove end slices=1]
		\lstick{\ket{0}}  & \gate{H} & \ctrl{1} & \qw & \qw \rstick[2]{$\frac{|{00}\rangle + |{11}\rangle}{\sqrt{2}}$}\\
	\lstick{\ket{0}}  & \qw      & \targ    & \qw & \qw & \qw
\end{quantikz}
   }
\caption{Quantum circuit with two time slices for the creation of an entangled state using a Hadamard gate followed by a CNOT gate~\cite{Nielsen.2011}.}
	\label{fig:entanglement}
\end{figure}
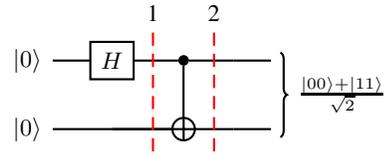
Entanglement refers to the phenomenon that measuring only one of the qubits determines the measurement outcome of the other qubit.
Mathematically, the \emph{Schmidt decomposition}~\cite{Schmidt.1908} theorem can be used to verify whether or not two qubits are entangled.
A quantum system described by its state vector $v$ is entangled if and only if the Schmidt rank of $v$ in its Schmidt decomposition is greater than one~\cite{Wilde.2013}.
Another type of gates are rotation gates.
Rotation gates can be used to rotate the state of a qubit by the angle $\theta$ around a specific axis of the Bloch Sphere.
Three commonly used rotation gates are $R_x(\theta), R_y(\theta)$ and $R_z(\theta)$ for the rotation around the $x$-, $y$- and $z$-axis.
A special type of rotation gate is the Pauli-$X$ gate which rotates a qubit's state exactly by $180^{\circ}$ around the $x$-axis.

\subsection{Patterns for quantum computing}
Patterns for quantum computing provide proven solutions to recurring problems that occur during the implementation process of a quantum algorithm~\cite{Leymann.2019}.
This contribution comprises algorithms for the automatic detection of eight quantum computational patterns.
The patterns selected are those that have a structure which is simple to detect and are commonly used.
Patterns that are not included in our tool are either more difficult to recognize or not widely used.
More detailed descriptions of the patterns can be found in the works of Leymann et al.~\cite{Leymann.2019} and Weigold et al.~\cite{ Weigold.2021, Weigold.2022}.
\begin{description}
    \item[Creating Entanglement:] This pattern describes the transition from an unentangled to an entangled state within a quantum algorithm.
    \item[Uncompute:] 
    Uncompute proposes a solution to remove unwanted entanglements using the quantum circuit proposed by Dervovic et al.~\cite{Dervovic.2018} shown in Fig.~\ref{fig:uncompute}.
\begin{figure}[tb]
	 \centering
     \resizebox{0.9\linewidth}{!}{ 
	    \begin{quantikz}[row sep=0.2cm]
	 	\lstick{$\ket{x}$} 	   & \qwbundle[alternate]{} & \qwbundle[alternate]{} & \qwbundle[alternate]{} &  \gate[7, bundle={1,2}, nwires={3,4,5,6,7}][1.2cm]{U_f^{-1}} \qwbundle[alternate]{} & \qwbundle[alternate]{} & \qwbundle[alternate]{} & \qwbundle[alternate]{} & \qwbundle[alternate]{} \rstick{\ket{x}}\\
		\lstick{$\ket{g(x)}$}  & \qwbundle[alternate]{} & \qwbundle[alternate]{} & \qwbundle[alternate]{} & \qwbundle[alternate]{} & \qwbundle[alternate]{} & \qwbundle[alternate]{} & \qwbundle[alternate]{} & \qwbundle[alternate]{} \rstick{\ket{0}} \\
	 	\\
		\lstick{$\ket{f_1(x)}$}   & \ctrl{4} & \qw & \qw & \qw & \swap{4} & \qw& \qw & \qw\rstick{\ket{f_1(x)}}\\
		\lstick{$\ket{f_2(x)}$}   & \qw & \ctrl{4} & \qw & \qw & \qw & \swap{4} & \qw & \qw\rstick{\ket{f_2(x)}}\\
		\lstick{$\vdots$}\phantom{\ket{f_1(x)}} & & & \vdots & & & & \vdots  & \phantom{\ket{f_11(x)}}\vdots\\
		\lstick{$\ket{f_m(x)}$}   & \qw & \qw &\ctrl{4} & \qw & \qw & \qw & \swap{4} & \qw\rstick{\ket{f_m(x)}}\\
		\lstick{$\ket{0}$}        & \targ{} & \qw &\qw &\qw & \targX{} & \qw & \qw & \qw\rstick{\ket{0}}\\
		\lstick{$\ket{0}$}        & \qw & \targ{}&\qw & \qw & \qw & \targX{} & \qw & \qw\rstick{\ket{0}} \\
		\lstick{$\vdots$}\phantom{\ket{0}} & & \vdots & & & & \vdots & & \phantom{\ket{0_{0l}}}\vdots \\
		\lstick{$\ket{0}$}        & \qw & \qw & \targ{} & \qw & \qw & \qw & \targX{} & \qw\rstick{\ket{0}} \\
	\end{quantikz}
    }
	\caption{Quantum circuit for the Uncompute procedure~\cite{Dervovic.2018}. $\ket{f_i(x)}$ refers to the qubit states in $\ket{f(x)}=\ket{f_1(x)f_2(x)\ldots f_m(x)}$. 
    }
	\label{fig:uncompute}
\end{figure}
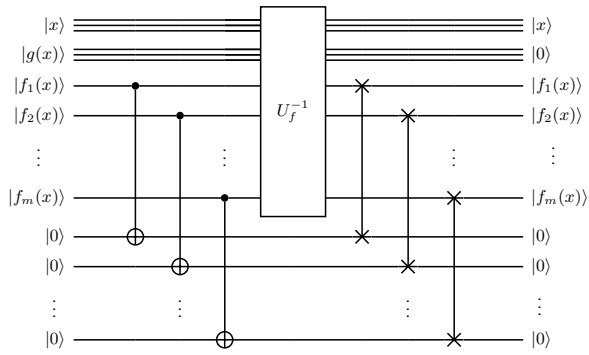
\end{description}
\begin{description}
    \item[Uniform Superposition:] This pattern is used to create a uniform superposition state by applying Hadamard gates to each qubit~\cite{Leymann.2019}.
    \item[Basis Encoding:] Classical information is converted into a quantum state by approximating the given input number in binary format and then encoding each bit into the state of a qubit using Pauli-$X$ gates~\cite{Weigold.2022}.
    \item[Angle Encoding:] This pattern encodes classical data by applying the rotation gate $R_y$, whereby the rotation angle is equal to the value of the normalized data point~\cite{Weigold.2021}.
    \item[Amplitude Encoding:] Another compact way of representing classical data is to encode the values into the amplitudes of the qubits~\cite{Weigold.2022}, which has been implemented by Shende et al.~\cite{Shende.2006} and others~\cite{Plesch.2011,Iten.2016,Mottonen.2005,Bergholm.2018}.
    \item[Quantum Phase Estimation:] Some quantum algorithms require the eigenvalue of a unitary transformation to be estimated~\cite{Weigold.2021}, which can be achieved using the quantum circuit described by Nielsen et al.~\cite{Nielsen.2011}.
    \item[Post Selective Measurement:] The continuation of a quantum algorithm is conditioned on a specific measurement result, allowing the algorithm to proceed if the desired result is obtained, or otherwise restart~\cite{Weigold.2021}.
\end{description}

\section{State- and Circuit-based Analysis for Quantum Pattern Detection}
\label{sec:contribution}
Quantum computing patterns can be characterized by their implementation at gate level and the way in which they transform the state of a quantum system.
For example, the pattern Basis Encoding is implemented using Pauli-$X$ gates in the first layer of the quantum circuit and transforms the quantum state by encoding a classical value into a quantum register.
Based on these two ways of characterization, we propose two basic approaches for recognizing a particular pattern.
The first one is to perform a static analysis by identifying special structures on gate level that are typical for a pattern.
These can be specific gate sequences or special gates that are very characteristic for the pattern.
We call detection algorithms, that use this approach, \emph{circuit-based} algorithms.
The second approach is to analyze the quantum state of the system during the execution of the algorithm using a dynamic code analysis.
With these states, we can derive knowledge about properties of the quantum system which can then be used to match the given code with a certain pattern.
We refer to algorithms that use this approach as \emph{state-based} algorithms.
In general, not every pattern that can be detected by a state-based approach can also be accurately identified with a circuit-based algorithm, and vice versa.
The reason for that is that most patterns are only consistent in one of the two properties.
For example, there are many ways to entangle a quantum state at gate level but the result of the Creating Entanglement pattern is always an entangled state.
In contrast to that, Basis Encoding is normally implemented using Pauli-$X$ gates but the resulting quantum state is always different depending on the encoded value.
Therefore, there is often only one approach that is suitable for detecting a certain pattern.
The pattern detection algorithms in our framework can be grouped according to these two approaches as shown in Tab.~\ref{table:overview}.
For each algorithm, we determined its theoretical time complexity.
\begin{table}[tb]
    \centering
    \begin{adjustbox}{width=\linewidth,center=\linewidth} 
    \begin{threeparttable}
    \begin{tabular}{l c c c} 
    	\textbf{Quantum Pattern} & \begin{tabular}[x]{@{}c@{}}\textbf{state-}\\\textbf{based}\end{tabular} & \begin{tabular}[x]{@{}c@{}}\textbf{circuit-}\\\textbf{based}\end{tabular} & \begin{tabular}[x]{@{}c@{}}\textbf{Theoretical}\\\textbf{Time-complexity}\end{tabular} \\ 
    	\toprule
    	Uniform Superposition (US) & \ding{55} &  & $\mathcal{O}(k\cdot 2^n)$ \\ 
    	Creating Entanglement (CE) & \ding{55} &  & $\mathcal{O}(k\cdot 2^n)$ \\
    	Basis Encoding (BE) & & \ding{55} & $\mathcal{O}(n)$ \\
        Angle Encoding (AE) & & \ding{55} & $\mathcal{O}(n)$ \\
    	Amplitude Encoding (AMP) & & \ding{55} & $\mathcal{O}(n\cdot m)$ \\
    	Quantum Phase Estimation (QPE) & & \ding{55} & $\mathcal{O}(n\cdot m)$ \\
    	Uncompute (UNC) & & \ding{55} & $\mathcal{O}(n\cdot m^4)$ \\
    	Post Selective Measurement (PSM) & & \ding{55}\tnote{*} & $\mathcal{O}(n\cdot m)$ \\
     \end{tabular}
    \begin{tablenotes}\scriptsize
        \item [*] The detector for Post Selective Measurement is not solely circuit-based since it also takes the implementation on code level into consideration.
    \end{tablenotes}
    \end{threeparttable}
    \end{adjustbox}
    \caption{Overview of properties of all pattern detectors, where $n$ is the number of qubits, $m$ is the number of layers in the quantum circuit and $k$ is the total number of unitary transformations applied within the system.}
    \label{table:overview}
\end{table}
In the following, we explain the detection approaches for the patterns Creating Entanglement and Uncompute in detail.
The implementation of all detection algorithms can be found on Github\footnote{\url{https://github.com/KIT-TVA/quantum-pattern-detector}}.


\subsection{Creating Entanglement (State-based analysis)}
To detect the creation of entanglement, we perform a state-based approach by analyzing the quantum state of the circuit after each unitary transformation. 
Our detection algorithm (see Alg.~\ref{alg:ent-alg}) uses the Schmidt decomposition theorem to identify entangled states.
\begin{algorithm}[tb]
	\SetKw{kwIn}{in}
	\SetKw{kwAnd}{and}
	\SetKw{kwOf}{of}
	\SetKw{kwBreak}{break}
	\SetAlgoLined
	\DontPrintSemicolon 
	pattern\_instances $\leftarrow\emptyset$\;
	\BlankLine
	\ForEach{instruction $\in$ quantum algorithm}{
		\ForEach{bipartition \kwOf current quantum system}{
			decomp $\leftarrow$ \begin{minipage}[t]{5cm}%
				Compute Schmidt decomposition for bipartition\;
			\end{minipage}\;
			\If{Schmidt rank of decomp $> 1$ \\
                    \kwAnd previous state not entangled}{
				pattern\_instances.add(\text{instruction\_number})\;
			\kwBreak\;
			}
		}
	}
	\BlankLine
\Return pattern\_instances\;
\caption{Detection of Creating Entanglement}
\label{alg:ent-alg}
\end{algorithm}
It divides the quantum system into every possible combination of two distinct subsystems and computes the Schmidt decomposition and Schmidt rank of the state vectors with respect to each bipartition (line 3-4).
If there exists one Schmidt rank that is greater than one, it can be concluded that the given state is entangled, otherwise, the state is not entangled.
If a change from an unentangled to an entangled state is detected, the algorithm returns this as an instance of the pattern Creating Entanglement (line 5-9).
This process is repeated for each program instruction in the given quantum algorithm (line 2).
The state vector at time slice 1 of the quantum circuit in Fig.~\ref{fig:entanglement} is $\frac{1}{\sqrt{2}}(\ket{00}+\ket{01})$.
The only Schmidt coefficient for this quantum state is 1 for every bipartition of the system. 
Thus, the quantum state is not entangled.
However, the Schmidt rank of the quantum state at time slice 2 is 2 with both Schmidt coefficients being $\frac{1}{\sqrt{2}}$ for the bipartition $\ket{q_0}\otimes\ket{q_1}$.
Therefore, our algorithm detects this as an instance of Creating Entanglement since there is a change from an unentangled to an entangled quantum state.
In line 3 of Alg.~\ref{alg:ent-alg}, the algorithm computes every bipartition of the quantum state.
Since the number of bipartite system divisions grows exponentially with the number of qubits, the total runtime complexity of Alg.~\ref{alg:ent-alg} is also exponential.
The advantage of this approach is that entanglement can be reliably detected using the Schmidt decomposition, making it suitable for smaller quantum systems.

\subsection{Uncompute (Circuit-based analysis)}
Uncompute is implemented on gate level using the quantum circuit shown in Fig.~\ref{fig:uncompute}~\cite{Dervovic.2018}.
This circuit can be divided into three parts.
In the first part, the state of the register $\ket{f(x)}$ is copied into the ancilla register by applying CNOT gates on each ancilla bit using the qubits in $\ket{f(x)}$ as control.
In the second part, $U_f^{-1}$ is applied on each register except the ancilla register.
In the last part, swap gates are used between the qubits in the ancilla register and the qubits in the register $\ket{f(x)}$.
Our circuit-based detection algorithm attempts to recognize this circuit structure for a given quantum algorithm.
It can be observed that in real implementations, the first and last part of the characteristic subcircuit are often omitted, i.e. copying the state into an ancilla register and restoring it with swap gates afterwards.
The reason for that is that the working register $\ket{g(x)}$ and the output register $\ket{f(x)}$ are often not entangled so that the garbage state can be reset without affecting the output.
Thus, copying the output beforehand becomes obsolete.
Although we also implemented an algorithm for detecting these two parts, the main focus is on detecting the inverse subcircuit $U_f^{-1}$.
\begin{algorithm}[tb]
 	\SetKw{kwIn}{in}
 	\SetKw{kwAnd}{and}
 	\SetKw{kwOf}{of}
	\SetKw{kwBreak}{break}
	\SetAlgoLined
	\DontPrintSemicolon 
	$m\leftarrow$ Number of layers in the quantum circuit\;
	\BlankLine
	\ForEach{$i$ \kwIn $\{1,\ldots,\lfloor m/2\rfloor\}$}{
	 	\ForEach{subcircuit of size $i$}{
             \begin{minipage}[t]{7cm}%
 				Check if there is a subsequent subcircuit of size $i$ that is inverse to the current subcircuit\;
 			\end{minipage}
 			\If{inverse subcircuit found}{
 				\Return{True}\;
	 		}
	 	}
	 }
 	\BlankLine
 	\Return{False}
	\caption{Detection of an inverse subcircuit}
 	\label{alg:inv}
\end{algorithm}
To detect the inverse subcircuit, every combination of two subsequent subcircuits with equal size is analyzed and it is verified if they are the inverse of each other (see Alg.~\ref{alg:inv}), for example by using the {\fontfamily{lmss}\selectfont  inverse}\footnote{\url{https://docs.quantum.ibm.com/api/qiskit/qiskit.circuit.QuantumCircuit\#inverse}} method from Qiskit~\cite{Treinish.2023}.
Let $n$ be the number of qubits and $m$ the number of layers in the underlying quantum circuit.
There is a maximum of $m$ subcircuits with size $i$ in every quantum circuit.
For each subcircuit, there are not more than $m$ subsequent subcircuits.
The comparison of each pair of subcircuits can be done in $\mathcal{O}(n\cdot m)$.
Therefore, the runtime of the inner loop in line 3 of Alg.~\ref{alg:inv} is in $\mathcal{O}(n\cdot m^3)$.
Since this procedure is repeated $\lfloor m/2 \rfloor$ times, an upper bound for the runtime of Alg.~\ref{alg:inv} is $\mathcal{O}(n\cdot m^4)$.
Due to the fact that not every occurrence of an inverse subcircuit belongs to a pattern instance, the detection algorithm can be further improved by specifying as a precondition that the state must have been previously entangled.

\section{Evaluation}
\label{sec:evaluation}
We evaluate our quantum pattern detection framework by investigating the following research questions:
\begin{description}
    \item[RQ 1 (Accuracy):] How correctly can our detection framework recognize patterns for quantum computing in terms of precision, recall and $\text{F}_1$-measure?
    \item[RQ 2 (Scalability):] How well do our detection algorithms scale with the sizes of the given quantum circuits?
    \item[RQ 3 (Comparison):] How does our framework compare with other existing detection frameworks for quantum computing patterns in terms of detection accuracy?
\end{description}

\subsection{Subject Systems and Ground Truth}
In order to address the research questions, we need a dataset with a ground truth against which we can evaluate our framework.
However, currently, there is no labeled dataset for benchmarking quantum pattern detection approaches.
We close this gap by creating a ground truth for subject systems selected from MQT Bench~\cite{Quetschlich.2023} and Qiskit~0.45.0~\cite{Treinish.2023} by manually determining the quantum patterns present in the underlying test code.
The subject systems were chosen from the code base of MQT Bench~\cite{Quetschlich.2023} and Qiskit~0.45.0~\cite{Treinish.2023} since they contain implementations of popular and widely used quantum algorithms, increasing the representativeness of our dataset.
Each quantum algorithm is represented by a Python function that, given some input parameters such as the number of qubits, builds a quantum circuit that implements the algorithm. 
These algorithms are then converted into OpenQASM code which serves as input for our detection programs during the evaluation process.
The Open Quantum Assembly Language (OpenQASM)~\cite{Cross.2017} is a low-level imperative programming language designed to describe quantum circuits and quantum algorithms.
We selected OpenQASM as the input format since it currently belongs to the de facto standards for hardware-independent exchange formats~\cite{Cross.2022b}.
To decide whether quantum patterns are present in the underlying subject systems, we use algorithm documentations and scan the source code manually for pattern occurrences. 
An overview of which pattern occurs in which algorithm is shown in Tab.~\ref{table:ground-truth}.
\begin{table}[tb]
    \centering
    \addtolength{\tabcolsep}{-0.3em}
    \resizebox{\linewidth}{!}{ 
    \begin{tabular}{l | c c c c c c c c} 
    	\multicolumn{1}{c |}{} & \multicolumn{8}{c}{\bfseries Ground Truth}\\
    	\bfseries Algorithm & US & CE & BE & AE & AMP & QPE & UNC & PSM \\ 
    	\midrule
    	Adder w. overflow & & & \ding{51} & & & & &\\
    	Adder w/o overflow & & & \ding{51} & & & & &\\
    	Amplitude Encoding & & \ding{51} & & & \ding{51} & & &\\
    	Amplitude Estimation & \ding{51} & \ding{51} & & & & \ding{51} & &\\
    	Deutsch-Jozsa & \ding{51} & & & & & & \ding{51} &\\
    	GHZ & \ding{51} & \ding{51} & & & & & &\\
    	Graph State & \ding{51} & \ding{51} & & & & & &\\
    	Grover & \ding{51} & \ding{51} & & & & & \ding{51} &\\
    	HHL & \ding{51} & \ding{51} & & & & \ding{51} & \ding{51} & \ding{51}\\
    	Multiplier & & & \ding{51} & & & & &\\
    	QAOA & \ding{51} & \ding{51} & & & & & &\\
    	QFT & \ding{51} & & & & & & &\\
    	QFT w. entanglement & \ding{51} & \ding{51} & & & & & &\\
    	OPE & \ding{51} & \ding{51} & & & & \ding{51} & &\\
    	Quantum Walk & \ding{51} & \ding{51} & & & & & \ding{51} &\\
    	Real Amplitudes & & \ding{51} & & \ding{51} & & & &\\
    	Shor & \ding{51} & \ding{51} & & & & \ding{51} & &\\
    	SU2 Ansatz  & & \ding{51} & & \ding{51} & & & &\\
    	VQE & & \ding{51} & & \ding{51} & & & &\\
    	W-State & & \ding{51} & & & & & &\\
    \end{tabular}
    }
    \caption{Abstraction of the ground truth. The symbol \ding{51} marks whether the corresponding patterns are present in each algorithm.}
    \label{table:ground-truth}
\end{table}
For the evaluation of the scalability of our detection programs, we execute our framework on randomly generated quantum circuits, each time with an increasing number of qubits and layers.
Finally, we use the dataset of Pérez-Castillo et al.~\cite{Perez.2024} to compare our framework with their detection method.
Since their dataset lacks ground truth, we also create a ground truth for a subset of their subject systems.


\subsection{Methodology}
We conduct four experiments during the evaluation, each of which addresses one specific research question.
In the first experiment, we execute our detection framework on all subject systems from Tab.~\ref{table:ground-truth} and compare the detection results with the ground truth which we have created.
As a result, we obtain a set of true positives (TP), false positives (FP), and false negatives (FN) for each subject system, which we can use to compute the evaluation metrics of precision, recall and $\text{F}_1$-measure~\cite{Powers.2011}.
The obtained metrics are then analyzed to provide an answer to RQ 1.

In the second experiment, we aim to make a statement about the scalability of our framework.
For that, we measure the runtimes of our framework for randomly generated quantum circuits with varying sizes.
The size of a quantum circuit is determined by the number of qubits $n$ (\emph{circuit width}) used within the algorithm and the number of layers $m$ (\emph{circuit depth}) in the quantum circuit.
These parameters can be specified independently for the creation of random circuits using the {\fontfamily{lmss}\selectfont  random\_circuit}\footnote{\url{https://docs.quantum.ibm.com/api/qiskit/0.45/circuit\#random\_circuit}} method from Qiskit. 
In order to analyze the scalability with respect to both input dimensions, we perform two measurement iterations in total.
In the first iteration, we set the number of layers $m$ in the circuit to a constant value ($m=5$) and measure the execution times depending on an increasing number of qubits.
In the second iteration, the number of qubits $n$ is fixed ($n=3$) and the number of layers is increased.
The constant values for $m$ and $n$ are chosen to be relatively small in order to minimize their impact on runtime. 
However, they should also not be selected so small that the resulting quantum circuit becomes trivial.
Furthermore, we have opted for values $m>n$, as quantum circuits generally have greater depth than width.
In each iteration, we repeat the runtime measurement 20 times for each detection algorithm and compute the average runtime values.
The benchmark system used in our experiments includes an Intel Core i5-10210U CPU with an integrated graphics processor and 8 GB RAM. 
Using this runtime information, we are able to draw conclusions about the scalability of each detection algorithm in our framework and answer RQ 2.

To answer RQ 3, we conduct two cross-validation experiments where we evaluate our framework against the only other known quantum pattern detection approach proposed by Pérez-Castillo et al.~\cite{Perez.2024}.
In the first cross-validation experiment, we execute our detection framework on the evaluation set created by Pérez-Castillo et al.~\cite{Perez.2024} and compare the detection results respectively.
The dataset of Pérez-Castillo et al.~\cite{Perez.2024} consists of 80 quantum circuits and is used for evaluating their detection program which is able to identify the patterns Uniform Superposition, Creating Entanglement, Uncompute, Initialization, and Oracle~\cite{Leymann.2019}.
In the second cross-validation experiment, we aim to make a conclusion about the detection accuracy of our and their pattern detection implementations.
To achieve this, we examine a subset of all subject systems and determine if the patterns Uniform Superposition and Creating Entanglement are present in these code fragments.
We only consider a subset of their subject systems due to the lack of documentation.
Many of the algorithms cannot be found since there is no comprehensible information regarding the source of the subject systems' implementations~\cite{Perez.data.2023}.
We have selected the patterns Uniform Superposition and Creating Entanglement since they can be detected using both our and their detection tool.
The third pattern that can be detected with both implementations is Uncompute.
However, since there is no documentation for the subject systems and most algorithms are not commonly used, it is challenging to identify intentional usages of the Uncompute pattern.
Using the information from the code examination, we calculate evaluation metrics for both approaches and draw conclusions about the accuracy of both detection implementations.

\subsection{Results}

In the first experiment, we measure the metrics precision, recall and $\text{F}_1$-measure after executing our detection framework on our set of subject systems. 
The results for each quantum pattern are shown in Tab.~\ref{table:er-ex1}.
\begin{table}[tb]
    \centering
    \resizebox{\linewidth}{!}{ 
    \begin{tabular}{l c c c} 
    	\textbf{Quantum Pattern} & \textbf{Precision} & \textbf{Recall} & \textbf{$\text{F}_1$-Measure} \\ 
    	\toprule
    	Uniform Superposition (US) & 1.0  & 1.0 & 1.0 \\ 
    	Creating Entanglement (CE) & 1.0 & 1.0 & 1.0 \\
    	Basis Encoding (BE) & 0.95 & 1.0 & 0.9744 \\
    	Angle Encoding (AE) & 0.85 & 1.0 & 0.9189 \\
    	Amplitude Encoding (AMP) & 0.95 & 1.0 & 0.9744 \\
    	Quantum Phase Estimation (QPE) & 0.8 & 1.0 & 0.8889 \\
    	Uncompute (UNC) & 0.75 & 1.0 & 0.8571 \\
    	Post Selective Measurement (PSM) & 1.0 & 1.0 & 1.0 \\
    \end{tabular}
    }
    \caption{Values for precision, recall and $\text{F}_1$-measure grouped by each quantum pattern in our benchmark set.}
    \label{table:er-ex1}
\end{table}
It can be observed that every quantum pattern can be detected with high precision values and a recall of 1 using our detection framework.
Furthermore, all state-based approaches achieve perfect results for precision, recall and $\text{F}_1$-measure.

In the second evaluation experiment, we address the question about the scalability of our detection programs.
The results are shown in Fig.~\ref{fig:execution-time}.
\pgfkeys{/pgf/number format/.cd,1000 sep={\,}}
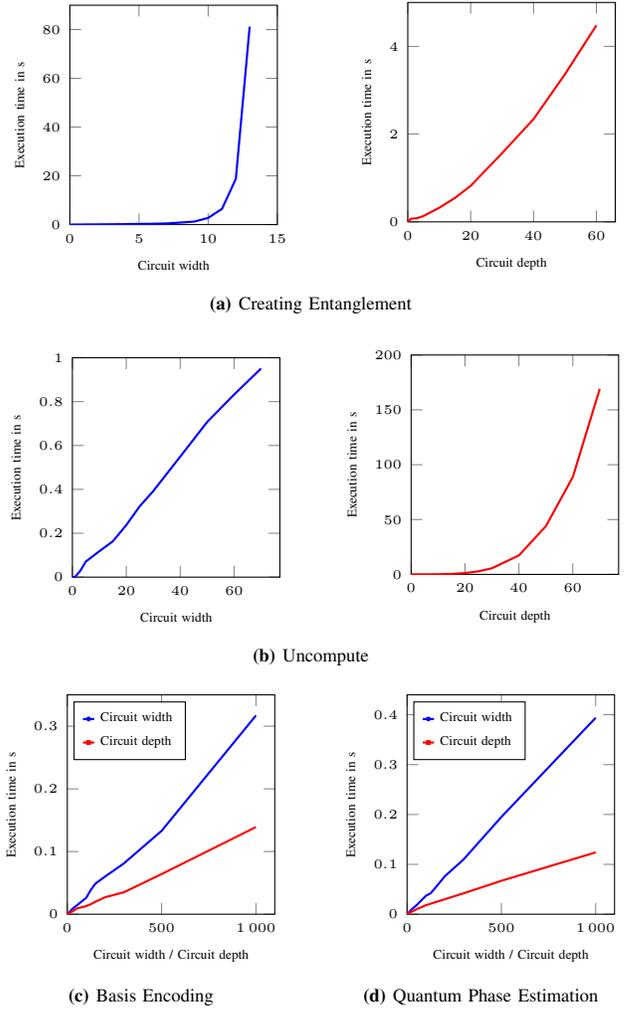
\begin{figure}[tb]
    \centering
    \begin{subfigure}[b]{\linewidth}
        \centering
        \begin{tikzpicture}
            \begin{axis}[
                width=0.49\linewidth,
                height=4.5cm,
                xmin=0,
                xmax=15,
                ymin=0,
                ymax=90,
                every axis plot/.append style={thick},
                tick label style={font=\tiny},
                xlabel={Circuit width},
                ylabel={Execution time in s},
                x label style={font=\tiny},
                y label style={font=\tiny}
                ]
                \addplot table [x=a, y=b, mark=none] {
                    a b
                    0 0
                    1 0.052
                    2 0.087
                    3 0.135
                    4 0.193
                    5 0.271
                    6 0.338
                    7 0.473
                    8 0.833
                    9 1.233
                    10 2.729
                    11 6.448
                    12 18.749
                    13 81.213
                };
            \end{axis}
        \end{tikzpicture}
        \hspace{1.5em}
        \centering
        \begin{tikzpicture}
            \begin{axis}[
                width=0.49\linewidth,
                height=4.5cm,
                xmin=0,
                ymin=0,
                ymax=5,
                yticklabel style={/pgf/number format/fixed, /pgf/number format/precision=3},
                every axis plot/.append style={thick},
                tick label style={font=\tiny},
                xlabel={Circuit depth},
                ylabel={Execution time in s},
                x label style={font=\tiny},
                y label style={font=\tiny}
                ]
                \addplot[red] table [x=a, y=b, mark=none] {
                    a b
                    0 0
                    1 0.071
                    3 0.084
                    5 0.133
                    10 0.321
                    15 0.544
                    20 0.821
                    25 1.194
                    30 1.565
                    40 2.341
                    50 3.365
                    60 4.479
                };
            \end{axis}
        \end{tikzpicture}
        \subcaption{Creating Entanglement}
        \label{fig:ce-et}
    \end{subfigure}

    \vspace{1em}

    \begin{subfigure}[b]{\linewidth}
        \centering
        \begin{tikzpicture}
            \begin{axis}[
                width=0.49\linewidth,
                height=4.5cm,
                xmin=0,
                ymin=0,
                ymax=1,
                yticklabel style={/pgf/number format/fixed, /pgf/number format/precision=3,
                every axis plot/.append style={thick}},
                tick label style={font=\tiny},
                xlabel={Circuit width},
                ylabel={Execution time in s},
                x label style={font=\tiny},
                y label style={font=\tiny}
                ]
                \addplot table [x=a, y=b, mark=none] {
                    a b
                    0 0
                    1 0.001
                    3 0.030
                    5 0.071
                    10 0.118
                    15 0.163
                    20 0.237
                    25 0.323
                    30 0.392
                    40 0.549
                    50 0.707
                    60 0.832
                    70 0.951
                };
            \end{axis}
        \end{tikzpicture}
        \hspace{1.5em}
        \centering
        \begin{tikzpicture}
            \begin{axis}[
                width=0.49\linewidth,
                height=4.5cm,
                xmin=0,
                ymin=0,
                ymax=200,
                yticklabel style={/pgf/number format/fixed, /pgf/number format/precision=3,
                every axis plot/.append style={thick}},
                tick label style={font=\tiny},
                xlabel={Circuit depth},
                ylabel={Execution time in s},
                x label style={font=\tiny},
                y label style={font=\tiny}
                ]
                \addplot[red] table [x=a, y=b, mark=none] {
                    a b
                    0 0
                    1 0.001
                    3 0.003
                    5 0.011
                    10 0.108
                    15 0.421
                    20 1.196
                    25 2.791
                    30 5.596
                    40 17.304
                    50 43.933
                    60 88.779
                    70 169.136
                };
            \end{axis}
        \end{tikzpicture}
        \caption{Uncompute}
        \label{fig:unc-et}
    \end{subfigure}

    \vspace{1em}

    \begin{subfigure}[b]{0.49\linewidth}
        \centering
        \begin{tikzpicture}
            \begin{axis}[
                width=\linewidth,
                height=4.5cm,
                xmin=0,
                ymin=0,
                ymax=0.35,
                yticklabel style={/pgf/number format/fixed, /pgf/number format/precision=3},
                every axis plot/.append style={thick},
                tick label style={font=\tiny},
                xlabel={Circuit width / Circuit depth},
                ylabel={Execution time in s},
                x label style={font=\tiny},
                y label style={font=\tiny},
                legend pos=north west,
                legend style={font=\tiny},
                legend cell align={left},
                legend image post style={scale=0.25}
                ]
                \addplot table [x=a, y=b, mark=none] {
                    a b
                    0 0
                    1 0.001
                    5 0.002
                    10 0.003
                    25 0.008
                    50 0.014
                    100 0.026
                    125 0.039
                    150 0.049
                    200 0.060
                    300 0.081
                    500 0.133
                    1000 0.317
                };
                \addlegendentry{Circuit width}

                \addplot table [x=a, y=b, mark=none] {
                    a b
                    0 0
                    1 0.001
                    5 0.001
                    10 0.002
                    25 0.004
                    50 0.009
                    100 0.013
                    125 0.016
                    150 0.020
                    200 0.027
                    300 0.035
                    500 0.064
                    1000 0.139
                };
                \addlegendentry{Circuit depth}
            \end{axis}
        \end{tikzpicture}
        \subcaption{Basis Encoding}
        \label{fig:be-et}
    \end{subfigure}
    \hfill
    \begin{subfigure}[b]{0.49\linewidth}
        \centering
        \begin{tikzpicture}
            \begin{axis}[
                width=\linewidth,
                height=4.5cm,
                xmin=0,
                ymin=0,
                ymax=0.44,
                yticklabel style={/pgf/number format/fixed, /pgf/number format/precision=3},
                every axis plot/.append style={thick},
                tick label style={font=\tiny},
                xlabel={Circuit width / Circuit depth},
                ylabel={Execution time in s},
                x label style={font=\tiny},
                y label style={font=\tiny},
                legend pos=north west,
                legend style={font=\tiny},
                legend cell align={left},
                legend image post style={scale=0.25}
                ]
                \addplot table [x=a, y=b, mark=none] {
                    a b
                    0 0
                    1 0.001
                    5 0.002
                    10 0.004
                    25 0.010
                    50 0.018
                    100 0.037
                    125 0.042
                    150 0.053
                    200 0.076
                    300 0.110
                    500 0.195
                    1000 0.394
                };
                \addlegendentry{Circuit width}

                \addplot table [x=a, y=b, mark=none] {
                    a b
                    0 0
                    1 0.001
                    5 0.001
                    10 0.003
                    25 0.005
                    50 0.010
                    100 0.018
                    125 0.021
                    150 0.024
                    200 0.030
                    300 0.042
                    500 0.067
                    1000 0.124
                };
                \addlegendentry{Circuit depth}
            \end{axis}
        \end{tikzpicture}
        \subcaption{Quantum Phase Estimation}
        \label{fig:qpe-et}
    \end{subfigure}
    \caption{Average execution times of different detection algorithms depending on the circuit width and depth.}
    \label{fig:execution-time}
\end{figure}
As illustrated in Tab.~\ref{table:overview}, the different
pattern detectors in our framework can be grouped into different classes of runtime complexity.
Accordingly, we have measured similar results for detectors of the same complexity class in our experiment.
As Fig.~\ref{fig:ce-et} indicates, the runtime development of all state-based pattern detectors are exponential in the number of qubits.
In the case of our generated subject systems, the runtime increases significantly after the number of qubits reaches 13.
However, Fig.~\ref{fig:ce-et} also shows that state-based pattern detectors scale with the number of layers in the quantum circuit since their execution times increase polynomially with the circuit depth.
The increase in execution times for all circuit-based pattern detectors in our framework are both polynomial in the number of qubits and layers in the quantum circuit.
For example, both detectors for Basis Encoding and Quantum Phase Estimation achieve execution times of less than 0.5 seconds on our benchmark system for an input size of 1000 qubits or layers (see Fig~\ref{fig:be-et} and Fig.~\ref{fig:qpe-et}) which is sufficient for practical use cases. 
Fig~\ref{fig:unc-et} illustrates that the Uncompute detector has a polynomial runtime in both input parameters.

In the first cross-validation experiment, we executed our detection framework on the dataset of Pérez-Castillo et al.~\cite{Perez.2024}.
Fig.~\ref{fig:comp} displays the total number of patterns found using either our detection framework or their detection algorithms.
\begin{figure}[tb]
    \centering
    \resizebox{\linewidth}{!}{ 
    \begin{tikzpicture}
        \begin{axis}[
            ybar,
            axis x line=bottom,
            axis y line=left,
            ylabel=Pattern count,
            ymin=0,
            ymax=60,
            symbolic x coords={US, CE, UNC, BE, AE, AMP, QPE, PSM, INI, OR},
            xtick={US, CE, UNC, BE, AE, AMP, QPE, PSM, INI, OR},
            xticklabel style={anchor=base,yshift=-\baselineskip},
            bar width=10pt,
            width=12cm,
            height=8cm,
            enlarge x limits=0.07,
            legend style={at={(0.5,-0.15)}, anchor=north, legend columns=-1},
            ]
            \addplot coordinates {
                (US, 57)
                (CE, 25)
                (UNC, 17)
                (BE, 2)
                (AE, 0)
                (AMP, 4)
                (QPE, 17)
                (PSM, 2)
                (INI, 0)
                (OR, 0)};
            \addplot coordinates {
                (US, 28)
                (CE, 0)
                (UNC, 0)
                (BE, 0)
                (AE, 0)
                (AMP, 0)
                (QPE, 0)
                (PSM, 0)
                (INI, 8)
                (OR, 3)};
            \legend{Our framework, Pérez-Castillo et al.}
            \draw[thick, dashed]
                    ($(axis cs:UNC, 0)!.5!(axis cs:BE, 0)$) --
                    ($(axis cs:UNC, 60)!.5!(axis cs:BE, 60)$);
        \end{axis}
    \end{tikzpicture}
    }
    \caption{Comparison of the total number of patterns detected between our framework and the implementation of Pérez-Castillo et al.~\cite{Perez.2024}. The results of the patterns that both approaches can detect are shown to the left of the dashed line, while those that can only be recognized by one implementation are shown to the right.
    The abbreviations INI and OR correspond to the patterns Initialization and Oracle~\cite{Leymann.2019}.
    }
    \label{fig:comp}
\end{figure}
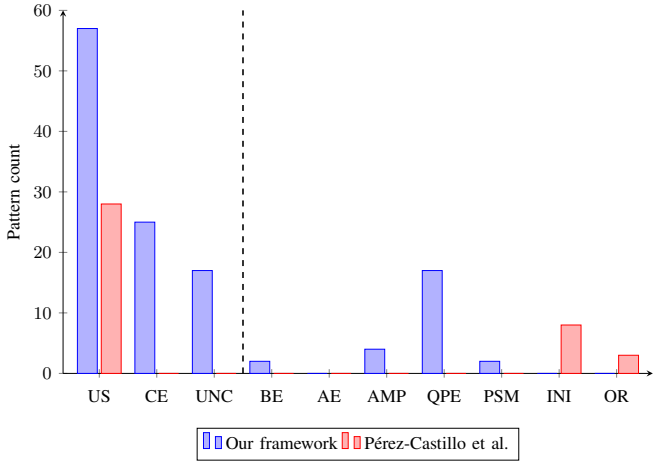
Our framework recognizes significantly more instances of patterns that can be identified by both implementations, in particular, it also detects instances of the patterns Creating Entanglement and Uncompute.
Furthermore, our framework is capable of detecting eight quantum computational patterns, while Pérez-Castillo et al.~\cite{Perez.2024} only offer implementations for five patterns.

The results of the second cross-validation experiment are displayed in Tab.~\ref{table:gt-perez}. 
It shows whether the pattern Uniform Superposition or Creating Entanglement occurs in 20 of the 80 subject systems from the dataset of Pérez-Castillo et al.~\cite{Perez.2024}.
Using this information, we calculate the precision and recall values for these two quantum patterns.
Our detection approaches achieve a precision and recall value of 1, whereas the detectors of Pérez-Castillo et al.~\cite{Perez.2024} also achieve a precision of 1 but only a recall of 0.31 for the Uniform Superposition detector.
Given that their approach does not recognize any instance of Creating Entanglement, it is not sensible to calculate a precision value for it. 
The recall value of their Creating Entanglement detector is 0.

\subsection{Discussion}
Using the evaluation results, we answer the research questions that were previously defined.

\textbf{RQ 1 (Accuracy):}
As shown in Tab.~\ref{table:er-ex1}, all state-based pattern detectors achieve perfect results regarding the detection accuracy, i.e. they achieve a $\text{F}_1$-measure of 1.
This is due to the fact that uniform superpositions and entanglements can be precisely identified by analyzing the state vector of the quantum system.
Both detectors do not use any heuristics or approximations for which reason they are absolutely reliable in terms of detecting the corresponding pattern.
Similarly, all circuit-based detectors achieve a recall of 1, as pattern-specific subcircuits are accurately detected.
However, since the circuit structures of these patterns may be used arbitrarily in a quantum algorithm without the actual intention of applying these patterns, the precision of these detectors is not perfect.
For Amplitude Encoding, we only considered one specific implementation proposed by Shende et al.~\cite{Shende.2006}, although there are many other possible implementations for this pattern~\cite{Plesch.2011,Iten.2016, Mottonen.2005, Bergholm.2018}.
If these were also taken into account for the ground truth, the recall value of our detector would be significantly lower.
Therefore, this detection algorithm could be expanded in subsequent work to detect other Amplitude Encoding techniques as well.
The detectors for Basis, Angle and Amplitude Encoding could be further improved by using machine learning approaches to determine the best threshold values used in these algorithms.
In summary, all of our detection algorithms achieve very high values for the $\text{F}_1$-measure.
Therefore, it can be concluded that quantum computational patterns can be detected with very high accuracy using our framework.
\begin{table}[tb]
    \centering
    \resizebox{\linewidth}{!}{ 
    \begin{tabular}{l c c c c} 
    	\begin{tabular}[x]{@{}c@{}}\textbf{Code fragment}\\\textbf{class name}\end{tabular} & \textbf{US} & \textbf{CE} & \begin{tabular}[x]{@{}c@{}}\textbf{Result of our}\\\textbf{framework}\end{tabular} & \textbf{Result of~\cite{Perez.2024}}\\ 
    	\toprule
    	python.teleport & \ding{51} & \ding{51} & US, CE & None\\
        test\_width\_pass & \ding{51} & & US & US \\
        test\_synthesis &  & \ding{51} & CE & None \\
        test\_dag\_to\_dagdependency & \ding{51} & & US & None \\
        test\_basic\_swap & \ding{51} & \ding{51} & US, CE & None \\
        test\_lookahead\_swap &  &  & None & None \\
        test\_dag\_fixed\_point\_pass & \ding{51} & \ding{51} & US,CE & None \\
        test\_resource\_estimation\_pass & \ding{51} &  & US & US \\
        circuits.teleport & \ding{51} & \ding{51} & US, CE & None \\
        example\_qiskit\_conditional & \ding{51} &  & US & US \\
        cnot\_logic &  &  & None & None \\
        qft\_4dec & \ding{51} &  & US & None \\
        fixed\_16 & \ding{51} & \ding{51} & US, CE & None \\
        qft\_3dec & \ding{51} &  & US & US \\
        buggy\_24 & \ding{51} &  & US & None \\
        test\_f16 & \ding{51} & \ding{51} & US, CE & None \\
        logic\_gates\_creator &  & & None & None \\
        quantum\_k\_means & \ding{51} & \ding{51} & US, CE & None \\
        qft\_3 & \ding{51} & & US & US \\
        swap & \ding{51} & \ding{51} & US, CE & None \\
     \end{tabular}
     }
    \caption{Overview of whether Uniform Superposition or Creating Entanglement occurs in the subject systems and whether it is recognized by the detection approaches. 
    }
    \label{table:gt-perez}
\end{table}

\textbf{RQ 2 (Scalability):}
As shown in Fig.~\ref{fig:execution-time}, the runtime development of all state-based detection algorithms is exponential in the number of qubits in the quantum circuit.
Therefore, the detectors for Uniform Superposition and Creating Entanglement both come with scalability problems regarding the circuit width. 
The reason for that is, that the detection algorithms perform an analysis over each state of the computational basis. 
The number of these quantum states grows exponentially with the size of the quantum circuit. 
This problem can potentially be solved by using heuristic detection approaches that consider only a subset of all possible quantum states in each iteration step.
Nevertheless, our state-based algorithms have a polynomial runtime complexity regarding the depth of the circuit.
Thus, they can be used for deep circuits with a limited width.
The increase in execution time for all circuit-based pattern detectors in our framework is polynomial. 
Hence, these detectors will scale with increasing hardware resources and are therefore feasible for very large quantum circuits.
The execution times of the Uncompute detector can also be further improved by providing more hardware resources due to its polynomial runtime complexity.
Furthermore, it can be noticed that the measured execution times for the Basis Encoding detector are not linear in only one input size parameter as indicated by the theoretical time complexity in Tab.~\ref{table:overview}.
The reason for that is that we use methods from Qiskit~\cite{Treinish.2023} for parsing and processing the quantum circuit.
Some of these methods like {\fontfamily{lmss}\selectfont  circuit\_to\_dag}\footnote{\url{https://github.com/Qiskit/qiskit/blob/stable/0.45/qiskit/converters/circuit\_to\_dag.py\#L19-L103}}, 
have a runtime linear in both circuit size parameters.
This runtime disparity is mainly caused by the use of Qiskit-specific implementations and could possibly be solved by using a different library or obtaining the required circuit data with a custom implementation.

\textbf{RQ 3 (Comparison):}
Comparing the evaluation results, Fig.~\ref{fig:comp} demonstrates that our framework detects significantly more of each pattern than the approach of Pérez-Castillo et al.~\cite{Perez.2024}.
The ground truth that we have created for the patterns Uniform Superposition and Creating Entanglement with respect to the subject systems of Pérez-Castillo et al.~\cite{Perez.2024} confirms that these patterns are detected correctly.
Furthermore, the detection accuracy of our framework for the Uncompute pattern has also to be higher, as this pattern is often used in conjunction with entanglements, making our detection result more sensible.
Unlike our framework, the detection algorithms of Pérez-Castillo et al.~\cite{Perez.2024} have not been tested for detection accuracy at all since their dataset lacks a ground truth.
Apart from that, it is unclear which type of instances of the Initialization pattern are recognized by the implementation of Pérez-Castillo et al.~\cite{Perez.2024}, as this pattern comprises several individual patterns such as Uniform Superposition or Basis Encoding according to the definition of Leymann~\cite{Leymann.2019}.
Thus, it can be concluded that our framework offers a more accurate detection approach than the approach of Pérez-Castillo et al.~\cite{Perez.2024}.
On top of that, our framework is capable of identifying a larger number of different patterns.

\subsection{Threats to Validity}

\subsubsection{Internal Validity}
It is possible that some patterns are labelled incorrectly in our benchmark set which can influence the results for precision and recall. 
To mitigate this issue, we confirmed the occurrences of specific patterns by frequently double-checking the created ground truth throughout the entire evaluation process.
By repeating the measurement multiple times and averaging the execution times, we tried to eliminate measurement inaccuracies on our benchmark system.

\subsubsection{External Validity}
The chosen subject systems may not reflect or contain enough patterns of interest to developers and researchers. 
This is due to the fact that the number of subject systems used in our evaluation is relatively small, e.g. there is only one quantum algorithm for Amplitude Encoding and Post Selective Measurement.
The best way to mitigate this risk is to further increase the set of subject systems in the future.
Another external threat is that nearly all quantum algorithm implementations, that are used for evaluation, were taken from MQT Bench~\cite{Quetschlich.2023}. 
It is possible that our detection programs overfit on these implementations since all quantum algorithms are implemented in a similar way. 
In order to mitigate this problem, we have confirmed the measurement outcomes for certain randomly chosen quantum algorithms on alternative implementations found on Github.
\section{Related Work}




The research community has extensively studied the detection and recovery of object-oriented design patterns~\cite{Gamma.1995}.
Multiple surveys have been published over time, such as by Dong et al.~\cite{Dong.2009} in 2009, by Rasool and Streitfdert~\cite{Rasool.2011} in 2011, by Al-Obeidallah et al.~\cite{AlObeidallah.2016} in 2016, and by Mzid et al.~\cite{Mzid.2024} in 2024. 
A ground truth and a standard benchmark set were missing in many of the works~\cite{Dong.2009, Rasool.2011, AlObeidallah.2016} and only a few works provided measured values for precision and recall~\cite{Dong.2009}.
Also, the exact locations of the detected patterns are often not provided~\cite{Rasool.2011} which complicates cross-validation against other approaches.
Many techniques only recover a few patterns that are relatively easy to detect, and often scalability and generalization are open questions~\cite{Rasool.2011}.
Most of these insights emphasize the need for a ground truth, that can be used to compare emerging pattern detection approaches.

In the quantum computing domain, static and dynamic code analysis have been used to improve the code quality and to identify bugs in quantum programs~\cite{Chen.2023,Paltenghi.2024,Zhao.2023}.
Xia et al.~\cite{Xia.2023} use static analysis to derive entanglement properties for Q\# programs via control flow graphs but they do not provide an implementation of their approach, leaving detection accuracy and scalability unevaluated~\cite{Xia.2023}.

{Jim{\'e}nez-Fern{\'a}ndez} et al.~\cite{Jimenez-Fernandez.2023} conducted a systematic mapping study on design patterns at the quantum circuit level.
The study mainly identified the pattern language proposed by Leyman et al.~\cite{Leymann.2019}.
The inital pattern language by Leyman et al.~\cite{Leymann.2019} was later extended several times with respect to data encoding~\cite{Weigold.2021, Weigold.2022}, hybrid quantum algorithms~\cite{Weigold.2021.2}, error handling~\cite{Beisel.2022}, and execution~\cite{Buehler.2023}.
The work of Khan et al.~\cite{Khan.2023} labeled high-level software architectures as patterns.
Nayak et al.~\cite{Nayak.2023} contributed a framework for the automatic detection of quantum bug-fix patterns using syntax trees and semantic checks.
However, they did not evaluate their framework on a dedicated set of subject systems.

To the best of our knowledge, the only available work on pattern detection for quantum software is by Pérez-Castillo et al.~\cite{Perez.2024}.
The authors employ a pattern detection technique based on a state machine.
From five considered patterns (Initialization, Uniform Superposition, Creating Entanglement, Oracle, and Uncompute), two patterns were not found at all during the evaluation, and the number of patterns that can be found with their tool is limited.
The dataset lacks a ground truth stating which pattern was expected to be found for each code fragment, which complicates a comparison.
\section{Conclusion and Future Work}
\label{sec:conclusion}
With this paper, we contributed a framework, consisting of eight individual algorithms, that is able to detect patterns for quantum computing automatically. 
Our detection algorithms are based on static and dynamic code analysis with both state-based and circuit-based approaches.
In the evaluation, we investigated on the accuracy and scalability of our framework and compared our framework with the only other known quantum pattern detection tool~\cite{Perez.2024} in terms of accuracy.
We showed that our state-based approaches achieve a perfect detection accuracy but do not scale well for larger input sizes.
In contrast to that, our circuit-based algorithms are scalable but less accurate than state-based approaches.
However, all of our detection algorithms are capable of identifying patterns very accurately.
Furthermore, we demonstrated the our framework outperforms the detection approach of Pérez-Castillo et al.~\cite{Perez.2024} in terms of detection accuracy.

In future work, we plan to extend our framework by developing and implementing more detection algorithms for quantum patterns that are currently not covered.
As explained in Sec.~\ref{sec:motivation}, the ultimate objective for the future would be to be able to use patterns as high-level building blocks for the construction of quantum software.
Our framework can assist in finding missing patterns by identifying code passages that are not covered by any pattern.
In order to achieve that, our framework has to be extended in such a way that it can output all code passages where a pattern has been used.

\balance
\bibliographystyle{IEEEtran}
\bibliography{IEEEabrv,bibliography}

\end{document}